\documentstyle[12pt]{article}

\begin{document}

\title{Nonperturbative Ground State \\
       of the Stochastic Stabilization \\
       of 2D Quantum Gravity}

\author{{\bf Oscar Diego}\thanks{e-mail: imtod67@cc.csic.es} \\
        {\em Instituto de Estructura de la Materia } \\
        {\em Serrano 123, 28006 Madrid} \\
        {\em Spain } }
\date{\mbox{}}

\maketitle

\thispagestyle{empty}

\begin{abstract}
I construct the ground state, up to first nonperturbative order,
of the stochastic stabilization of the
zero dimensional matrix model which defines 2D Quantum Gravity.
It is given by the linear combination of a perturbative wave function
and a nonperturbative one.
The nonperturbative behaviour which
arise from the stabilized model and from the string equation are
similar. I show the
modification of the loop equation by nonperturbative
contribution.
\end{abstract}

\begin{flushright}

\vspace{-13.5 cm} {IEM-FT-80/93}

\end{flushright}

\baselineskip=21pt

\vfill

\newpage
\setcounter{page}{1}

{\bf 1.	Introduction}

The zero-dimensional one matrix model leads to a nonperturbative
definition of simplicial 2D quantum gravity
coupled to conformal matter with $ c < 1 $\cite{dis,kaz,dou}.
In the continuum limit, the perturbative topological expansion in
the genus can be evaluated by solving
some differential equation: the string equation\cite{dou}.
The Schwinger-Dyson equations of the matrix model defines the
loop equation of the 2D quantum gravity\cite{kaz}.
The matrix model defines also the
nonperturbative behaviour up to the
ambiguity of the boundary condition of the
string equation. Unfortunately, in the case of pure
gravity, the nonperturbative real solutions of the string equation
for every boundary condition, can not be
solutions of the loop equation\cite{dal}.

The stochastic stabilization of the zero-dimensional
matrix model is defined by a one-dimensional matrix
model with the same perturbative expansion in $1/N$,
and perhaps it provides a consistent definition
of nonperturbative effects\cite{3,par}. In reference\cite{mig}
the scaling part of the stabilized hamiltoninan is solved
numerically. In this paper I propose an analytic approach in order
to show the relationship with the nonperturbative behaviour of the
original matrix model.

In section two, I review some results about the
nonperturbative effects of the matrix model in order
to compare it with the stabilized model.

In section three, the stochastic stabilization of the
matrix model is introduced, the relationship between the
eigenvalue density of the matrix model and the fermionic
density of the stabilized model, and between the
Schwinger-Dyson equation of the matrix model and the Ward
identities of the stabilized model are showed.

In section four, I perform the calculation of the nonperturbative
contribution in the case of pure gravity. This
calculation is performed using the old WKB method.
Finally, some conclusions are presented.

{\bf 2.	Nonperturbative 2D quantum gravity}

In this section I review some results of references\cite{zinn,dav}.

The nonperturbative effects of
the zero-dimensional matrix model are given by
the tunnelling of one eigenvalue
outside the main well of the potential.
The potential of pure gravity is
unbounded from below and the tunnelling takes place
between the well and the unbounded region.

In the large $N$ limit, the free energy is
\begin{equation}
\ln Z  = N^{2} \left ( \int d\lambda d\mu \rho (\lambda)
\rho (\mu) \ln\mid\lambda-\mu\mid
- \int d\lambda \rho (\lambda) W(\lambda) \right )
\end{equation}
where $\rho (\lambda)$ is the
eigenvalue density and $W(\lambda)$ is the
potential. The variation of the
free energy when one eigenvalue is
moved outside the support of $\rho
(\lambda)$ is
\begin{equation}
\Gamma (\lambda_{f}) = \delta \ln Z = N^{2} \int d \lambda
\delta \rho (\lambda) \left ( 2\int d\mu
\rho (\mu) \ln\mid\lambda-\mu\mid - W(\lambda) \right)
\label{eq:11}
\end{equation}
where $\delta \rho (\lambda) = N^{-1}
\{\delta (\lambda - \lambda_{f}) - \delta
(\lambda - \lambda_{i})\}$.
One can interpret $\Gamma (\lambda_{f})$ as the
effective potential for one eigenvalue
in the background created by the $N - 1$
remaining eigenvalues\cite{dav}. The formula (\ref{eq:11})
can be written\cite{zinn,bre}
\begin{eqnarray}
\Gamma (\lambda_{f}) & = & N
\int_{\lambda_{i}}^{\lambda_{f}} d \lambda
\left ( 2 \int d \mu \frac{\rho (\mu) }
{\lambda - \mu} - W^{\prime}
(\lambda) \right) \nonumber \\
  & = & - 2 N \int_{\lambda_{i}}^
{\lambda_{f}} d \lambda
U (\lambda) \sqrt{\lambda^{2} - a^{2}}
\end{eqnarray}
where $a$ is the cut of the support of $\rho(\lambda)$
and $U(\lambda)$ is a polynomial given by
the saddle point aproximation.
The maxima of the effective potential $\Gamma (\lambda_{f})$
are given by the
zeros of $U(\lambda)$. One may consider the instanton
configuration where one
eigenvalue sits at the top of $\Gamma(\lambda_{f})$. Taken
into account that the eigenvalue density is\cite{bre}
\begin{equation}
\rho (\lambda) = U(\lambda) \sqrt{a^{2} -
\lambda^{2}} \ , \ \lambda \in (-a,a)
\label{eq:6}
\end{equation}
one can write the instantonic contribution
as an integral of the imaginary
part of the analytic continuation of the eigenvalue density
\begin{equation}
\Gamma_{matrix} = - 2 N \int_{a}^{b} d\lambda \Im (\rho (\lambda))
\label{eq:12}
\end{equation}
where $a$ is the cut of the support of $\rho(\lambda)$
and $b$ is the top of the effective potential $\Gamma(\lambda_{f})$.
In the double scaling limit this
contribution is finite and is
the source of all the troubles of pure
quantum gravity. This nonperturbative contribution
arise also from the string equation\cite{zinn}.

The instantonic action (\ref{eq:11}) can be interperted as an
effective potential as follows: a source coupled to an eigenvalue
is added to the original potential
\begin{equation}
\sum_{i=1}^{N} W(\lambda_{i}) \longrightarrow
\sum_{i=1}^{N} W(\lambda_{i}) - J \lambda_{N}
\end{equation}
At leading order the source term only affect the eigenvalue
$\lambda_{N}$, so that the eigenvalue density is now
\begin{equation}
\tilde{\rho}(\lambda) = \rho (\lambda) +
\frac{1}{N} \{ \delta (\lambda -
\tilde{\lambda}_{N}) - \delta (\lambda - \bar{\lambda}_{N}) \}
\end{equation}
where $\tilde{\lambda}_{N}$ is the $N$ eigenvalue of the saddle point
configuration in the presence of the source and $\bar{\lambda}_{N}$
is the eigenvalue without the source. In the large $N$ limit
the classical field $\langle \lambda_{N} \rangle$
is given by the eigenvalue $\lambda_{N}$.
Hence, the effective potential
\begin{equation}
\Gamma ( \langle \lambda_{N} \rangle ) =
\ln Z(J) - \langle J \lambda_{N} \rangle
\end{equation}
becomes the instantonic action (\ref{eq:11}) when the classical field
and $\lambda_{f}$ are identified.

It is well know that $SU(N)$ symmetry is broken if there exist non
trivial solution of
\begin{equation}
\frac{\delta \Gamma}{\delta \lambda_{N}} = 0.
\end{equation}
Henceforth, the instantonic configurations with an eigenvalue at the
top of the effective potential break the $SU(N)$ symmetry.

Therefore, $SU(N)$ symmetry is restored like follow:
there are $N$ vacua $ |i \rangle $ with an eigenvalue $\lambda_{i}$
at the top of the
effective potential. SU(N) vacua are given by linear
combination of $ |i \rangle $ which must
be invariant under permutation of the
$N$ eigenvalues. There are only one linear combination and does not exist
arbitrary parameter like the $\theta $ parameter of gauge theories.

{\bf 3.	Stochastic stabilization of the zero-dimensional matrix model}

Stochastic stabilization\cite{3,par} provides a
nonperturbative definition of
2D quantum gravity, while reproducing
the perturbative expansion in powers of $ 1/N $ of
the original matrix model
and, therefore, the topological expansion of the
discretized quantum gravity.

The stochastic stabilization introduces a positive definite
hamiltonian
\begin{equation}
H = \frac{1}{2} Tr \left\{- \frac{1}{N^2}
\frac{\partial^2}{\partial\Phi^2} +
\frac{1}{4} \left( \frac{ \partial W }
{ \partial \Phi} \right)^{2} -
\frac{1}{2 N} \frac{\partial^2 W}{\partial\Phi^{2}}  \right\}
\label{eq:2}
\end{equation}
this hamiltonian is well defined
for all values of $N$ and coupling
constants.
The zero mode of the stabilized theory is
\begin{equation}
\Psi(\Phi) \sim \exp \left\{ - N \frac{W(\Phi)}{2} \right\}
\end{equation}
its norm being the partition
function of the original matrix model.
Hence, the matrix model
is well defined if and only if this
zero energy state is a
normalizable state. When this is the case,
corresponding obsevables in both
theories coincide
\begin{equation}
\langle Q \rangle_{stab} = \int Q \mid\Psi\mid^2 d\Phi =
\frac{1}{Z} \int Q \exp\Bigl\{- N W (\Phi)\Bigr\} d\Phi =
\langle Q \rangle_{matrix}
\label{eq:o1}
\end{equation}

In the ${\lambda_{n}}$ variables, where $\{\lambda_{n}\}$
are the eigenvalues
of the matrix $\Phi$, the zero energy state is the Slater determinant
\begin{equation}
\Psi (\{\lambda_{n}\}) = \prod_{i < k}^{N} (\lambda_{i} - \lambda_{j})
\exp \left\{ - \frac{N}{2} \sum_{n = 1}^{N} W(\lambda_{n}) \right\}
\label{eq:3}
\end{equation}
and, hence, the stabilized model is a Fermi gas.

In the planar limit, the condition (\ref{eq:o1}) becomes
\begin{equation}
\int_{a}^{b} d \lambda \lambda^{n} \rho_{matrix} =
\int_{a}^{b} d \lambda \lambda^{n} \rho_{stabilized}
\end{equation}
where $\rho_{matrix}$ and $\rho_{stabilized}$ are the density of eigenvalues
of the zero dimensional matrix
model and the fermionic density of the stabilized model. And taking into
account that both densities are the square root of a polynomial which is
zero at the
points $a$ and $b$ it is not difficult to prove that both densities
must be equal\cite{mir}.

The fermionic density in the
large $N$ limit for the stabilized hamiltonian is
\begin{equation}
\rho (\lambda) = \frac{1}{\pi} \sqrt{2(E_{F} - V)}
\label{eq:5}
\end{equation}
where $V$ is the stabilized potential\footnote{in the large $N$ limit, $V$ is a
one particle potential given by the Hartree aproximation} and $E_{F}$ is the
Fermi level. From the equality between (\ref{eq:5}) and (\ref{eq:6}),
$E_{F} - V$ must has a simple zero for $\lambda = a$ and
double zeros corresponding to zeros of $U(\lambda)$, in particular, a
double zero for $\lambda = b$. Therefore, in the
simplest case of a cuartic potential\cite{voz},
the stabilized model is given by a Fermi gas placed at the main well of the
stabilized potential, where the classical turning point of the Fermi level is
given by the cut of the eigenvalue density of the matrix
model: $a$. The stabilized potential has also a local minimun $b$,
which is reached by
the Fermi level, and it is given by the top of the effective potential of the
matrix model.
Hence, the main well of the stabilized potential defines the
perturbative expansion of the matrix model and I will show
how the local minimun defines the nonperturbative effects.

The stabilized hamiltonian (\ref{eq:2})
is the supersymmetric hamiltonian of
reference\cite{par} restricted to the bosonic sector.
The Schwinger-Dyson
equation of the zero-dimensional
matrix model becomes the Ward identities of
the supersymmetric matrix model.
The Ward identities of the stabilized
hamiltonian are given by the condition\cite{osc}
\begin{equation}
\frac{\partial E}{\partial g_{n}} = 0
\end{equation}
where $g_{n}$ are the coupling
constants of the zero-dimensional matrix model
and $E$ is the ground energy of the stabilized hamiltonian.
In appendix $2$, I extract from this condition the Virasoro
constraints.

{\bf 4.	Nonperturbative stochastic pure gravity}

Let us consider the matrix model
\begin{equation}
Z = \int d \Phi\, \exp \Bigl\{ - N Tr W ( \Phi ) \Bigr\}
\label{eq:20}
\end{equation}
where
\begin{equation}
W(\Phi) = Tr \Phi^2 - \frac{g}{2} Tr \Phi^4,
\end{equation}
and $\Phi$ is an $N$
dimensional hermitian matrix. This model corresponds to
a discretized formulation of pure quantum gravity\cite{dou}.

The Fokker-Planck hamiltonian of the stabilized model is the sum of $N$ one
particle hamiltonians\cite{voz}
\begin{equation}
h_{n} = - \frac{1}{2} \frac{1}{N^2} \frac{\partial^2}{\partial\lambda_n^2} +
\frac{1}{2} \{ g^2 \lambda_n^6 - 2g\lambda_n^4 + ( 1 + 2g ) \lambda_n^2
- 1 \}
\label{eq:21}
\end{equation}
and an interacting term
\begin{equation}
\frac{1}{2} \{ \frac{g}{N} \sum_{n,m} \lambda_n \lambda_m \}
\label{eq:A}
\end{equation}
where $\{\lambda_{n}\}$ are the eigenvalues of the matrix $\Phi$. The
interacting term is subleading in the $1/N$ expansion and I discard them.
The one particle stabilized potential (\ref{eq:21}) has a main well
and a local minimun below some critical coupling constant
and only one well above it.

The $1/N$ expansion is equivalent
to the WKB aproximation, and the quantization
condition is\cite{par}
\begin{equation}
\frac{N}{\pi} \int d\lambda \sqrt{2(E_{n}-V)} + \cdots = n + \frac{1}{2}
\label{eq:22}
\end{equation}
The Fermi level $E_{N-1}$ in the large $N$ limit is the value of
the one particle potential (\ref{eq:21}) in
its local minimun\cite{voz}.
The first correction to the Fermi
level is negative (appendix $3$). Henceforth,
in the perturbative expansion
$1/N$, the quantization condition
(\ref{eq:22}) restricted to the main well gives the correct perturbative
quantization of energy levels.

However, the perturbative WKB wave function has a
singularity in the local minimun.
It is well know that the WKB aproximation break down near
the turning points where the variation of the local wavelength is not small
\begin{equation}
\left | \frac{1}{2} \frac{d}{dx} \left (\frac{\hbar}{\sqrt{2(E-V)}} \right )
\right |   \sim 1
\end{equation}
At those points one needs the solution of
the Schr\"{o}dinger equation in order
to give the connection formulas and the quantization condition.

The local wavelength near the local minimun of the stabilized potential
(\ref{eq:21}) is
\begin{equation}
\Lambda (x) \sim \frac{\hbar}{\sqrt{2(e_{1} \hbar + x^{2})}}
\label{eq:23}
\end{equation}
where $e_{1} \hbar$ is the perturbative shift of the Fermi level and
$e_{1} > 0$ because the Fermi level is placed below the local minimun.
The origin of
coordinates is placed at the local minimun and $x = \lambda - b$.

The variation of the local wavelength (\ref{eq:23}) is order one
when $x$ is order $\sqrt{\hbar}$. Henceforth,
I perform the change of variables
\begin{equation}
y = \frac{x}{\sqrt{\hbar}},
\end{equation}
and the
Schr\"{o}dinger equation at first order in
$\hbar$ becomes
\begin{equation}
\frac{d^{2}}{dy^{2}} \Psi - ( e_{1} + C^{2} y^{2} ) \Psi = 0
\label{eq:24}
\end{equation}
In appendix $1$, I show the solutions of this equation and the
connection formulas in the local minimun which arise from them.

The more general wave function in the
main well of the stabilized potential is the
oscillatory solution
\begin{equation}
\Psi ( \lambda ) = \frac{A}{(2(E-V))^{ \frac{1}{4} } } \cos \left( N
\int_{0}^{\lambda} d x \sqrt{2(E-V)} + \delta \right ).
\end{equation}
Because the potential is even, one can
construct a wave function with definite
parity. For the even wave function
$\delta = 0$ and $\delta = \frac{\pi}{2}$
for the odd wave function.

The energy level may be written as the sum of a perturbative and
a nonperturbative term : $ E = e_{n} + \tilde{e}_{n} $.
Hence, the wave function becomes $ \Psi ( \lambda) =
\Psi^{p} (\lambda) + \Psi^{np} (\lambda) $, where

\begin{eqnarray}
\Psi^{np} (\lambda) & = & \frac{ - A N \tilde{e}_{n}
}{ 2 \omega_{n} (2(E-V))^{\frac{1}{4}}}
\sin{ \left \{ N \int_{\lambda}^{a} dx \sqrt{2(E-V)} - \frac{\pi}{4} + \eta
\right \}} \nonumber \\
\Psi^{p} (\lambda) & = & \frac{A}{(2(E-V))^{\frac{1}{4}}}
\cos{ \left \{ N \int_{\lambda}^{a} dx \sqrt{2(E-V)} - \frac{\pi}{4} + \eta
 \right \}}.
\end{eqnarray}
where
\begin{equation}
\eta = \frac{\pi}{4} - N \int_{0}^{a} dx \sqrt{2(e_{n}-V)} - \delta,
\end{equation}
$a$ is the classical turning point
and $\omega_{n}$ is the classical frecuency. The wave
amplitude of $\Psi^{np}(\lambda)$ is suppressed by the nonperturbative
term $\tilde{e}_{n}$, hence $\Psi^{p}(\lambda)$ is the perturbative part of
the wave function.

Beyond the local minimun the wave function must be the fall off
exponential in the infinity
\begin{equation}
\Psi (\lambda) = \frac{B}{(2(V-E))^{\frac{1}{4}}} \exp \left \{ -
N\int_{b}^{\lambda} dx \sqrt{2(V-E)} \right \}
\end{equation}
where $b$ is the local minimun.

The connection formula at the local minimun gives the wave function between the
local minimun and the main well (appendix $1$)
\begin{eqnarray}
\Psi (\lambda) \rightarrow \Psi (\lambda) & = & \frac{ B f_{1}
(e_{1},x_{0})}{(2(E-V))^{\frac{1}{4}}} \exp {\left \{ - N
\int^{b}_{\lambda} d x
\sqrt{2(V-E)} \right \}} \nonumber \\
 & + & \frac{ B f_{2}
(e_{1},x_{0})}{(2(E-V))^{\frac{1}{4}}} \exp {\left \{ + N
\int^{b}_{\lambda} dx
\sqrt{2(V-E)} \right \}}
\end{eqnarray}
where $x_{0}$ is an arbitrary constant. This wave function may be written as
$ \Psi (\lambda) = \Psi_{+} (\lambda) + \Psi_{-} (\lambda)$ where
\begin{eqnarray}
\frac{\Psi_{+}(\lambda)}{B f_{1}(e_{1},x_{0})} &  = &
\frac{ \tilde{A}}{(2(V-E))^{\frac{1}{4}}}
\exp{ \left \{ + N \int_{a}^{\lambda} dx
\sqrt{2(V-E)} \right \}} \nonumber \\
\frac{\Psi_{-}(\lambda)}{B f_{2}(e_{1},x_{0})} &  = &
\frac{\tilde{A}^{-1}}{(2(V-E))^{\frac{1}{4}}}
\exp{ \left \{ - N \int_{a}^{\lambda} dx \sqrt{2(V-E)} \right \}} \nonumber \\
\tilde{A} & = & \exp \left \{ - N \int_{a}^{b}dx \sqrt{2(V-E)} \right \}
\end{eqnarray}
where $\Psi_{+} (\lambda)$ is suppresed by a nonperturbative term, hence the
perturbative part of the wave function is $\Psi_{-}(\lambda)$.

The usual matching condition in the classical turning point $a$ between the
perturbative wave functions
$\Psi^{p}(\lambda)$ and $\Psi_{-}(\lambda)$ gives the
perturbative quatization condition (\ref{eq:22}) and the following relation
between the arbitrary constants $A$ and $B$
\begin{equation}
A = 2 B f_{2}(e_{1},x_{0}) \exp{ \left \{ N \int_{a}^{b} dx \sqrt{2(V-E)}
\right \} }
\end{equation}

And the matching condition between
$\Psi^{np}(\lambda)$ and $\Psi_{+}(\lambda)$ gives the
following value for the nonperturbative part of the energy level
\begin{equation}
\tilde{e}_{n} = \pm \frac{1}{N} \frac{f(e_{1},x_{0})}{\int_{a}^{b}
\frac{dx}{\sqrt{2(V-e_{n})}}} \exp{\left\{ - 2 N \int_{a}^{b} dx
\sqrt{2(V-e_{n})} \right \}}
\end{equation}
where, $f(e_{1},x_{0}) \sim x_{0} $ for $x_{0}$ small (appendix 1)
. This nonperturbative
corrections arise only for levels near the local minimun.

Therefore, the total energy of the ground state is
\begin{equation}
E = \sum_{n=0}^{N-1} \tilde{e}_{n}
\label{eq:26}
\end{equation}
where $\tilde{e}_{n}$ is nonzero only for levels near the Fermi energy
where
\begin{equation}
e_{n} = E_{F} + O(1/N)
\end{equation}
hence, in the large $N$ limit, (\ref{eq:26}) is approached by
\begin{equation}
E = \frac{x_{0}}{N} \exp \left ( -2N \int_{a}^{b} d \lambda \sqrt{2(V-E_{F})}
\right ) \sum_{n=0}^{N-1} \frac{1}{\int_{a}^{b}d \lambda \sqrt{2(V-e_{n})}}
\end{equation}

The nonperturbative contribution to the observables can be written as
\begin{equation}
K(g)\exp (\Gamma_{stabilized}),
\end{equation}
where the instantonic action is given by
\begin{equation}
\Gamma_{stabilized} = - 2 N \int_{a}^{b} d \lambda \Im (\rho(\lambda))
\label{eq:25}
\end{equation}
where $\rho(\lambda)$ is the analytic continuation of the
fermionic density of the stabilized model which is equal to the eigenvalue
density of the zero-dimensional matrix model.
Hence, the instantonic action of the matrix
model (\ref{eq:12}) and of the
stabilized model (\ref{eq:25}) are equal.
(\ref{eq:25}) also agree with the instantonic action
of reference\cite{mir}, which is given by a sucession of
instanton-antiinstanton starting and ending at the main
well.

The constant $x_{0}$ is analogous to the arbitrary constant of the
nonperturbative contribution which arise from the string equation
\cite{dou}. However, in the stabilized model there is
not a relationship between $x_{0}$ and boundary conditions
of some string equation, in fact the ambiguity arise
form the connection formulas (appendix 1).
In reference\cite{mig} the stabilized model is solved
numerically and the solution is unique, so the ambiguity
which I have found must be a defect of the approach, but
the arbitrary constant $x_{0}$ must be different from zero
because otherwise the WKB wave function is not a solution
of the Schr\"{o}dinger equation near the local minimun.

The important point is that the ground state energy is greater than zero by
a nonperturbative correction. From reference\cite{osc}
and (\ref{eq:26}) the first loop equation becomes in the double
scaling limit
\begin{equation}
\dot {V} \left(\frac{\partial}{\partial L } \right) \langle W (L) \rangle -
\int_{0}^{L} dJ \langle W(J) W( L - J ) \rangle \sim f(z) \exp \left \{ -
\frac{2}{5} \sqrt{6} z^{\frac{5}{2}} \right \} ( 1 + \cdots )
\label{eq:28}
\end{equation}
hence, the nonperturbative behaviour of this new loop equation is
compatible with the string equation of the zero dimensional matrix model.
But, in the stabilized model the string equation and the KdV flow
are lost because the Virasoro constraints also changes by nonperturbative
corrections. Is an open problem
what replaces the old string equation and the KdV flow.

{\bf 5. Conclusions}

I have found the first nonperturbative correction to the
ground state of the stochastic stabilization of zero
dimensional matrix model. The nonperturbative
behaviour of the original matrix model and its
stabilization are similar.

The loop equation of the stabilized
model and the zero-dimensional matrix model
differ by a nonperturbative term (\ref{eq:28}), which
is similar to the nonperturbative
behaviour of the string equation. Therefore, I expect
that the stabilized model gives a consistent
definition of the simplicial 2D quantum gravity.

  From the point of view of the
original matrix model, the nonperturbative
effects arise from the tunnelling
between the well of the potential to the
unbounded region. Hence, the perturbative
vacuum state is a metastable state
and it decay to an ill defined vacuum.
But, from the point of view of the
stabilized model there is only one
perturbative vacuum state and there is not
an ill defined vacuum\cite{osc,mig}. In ordinary quantum mechanics the
nonperturbative effects arise from tunnelling between two or more perturbative
vacua, the true nonperturbative vacuum is given by some
linear combination of perturbative vacua, and the physical interpretation is
that one perturbative vacuum decay by quantum tunnelling into the others.
However, the ground state of the stablized model is given
by the sum of a perturbative part, which is the perturbative wave function
around the main well, and a nonperturbative term
\begin{equation}
\Psi (\lambda) = \Psi^{p}(\lambda) + \Psi^{np}(\lambda).
\end{equation}

The perturbative wave
function is given by an oscillatory solution at the main well and a fall off
exponential at the classical forbidden region
\begin{eqnarray}
\Psi^{p} (\lambda) & = & \frac{A}{(2(E-V))^{\frac{1}{4}}}
\cos{ \left \{ N \int_{\lambda}^{a} dx \sqrt{2(E-V)} - \frac{\pi}{4} + \eta
 \right \}} \ \lambda < a  \nonumber \\
\Psi^{p}(\lambda) & = &
\frac{A}{(2(V-E))^{\frac{1}{4}}}
\exp{ \left \{ - N \int_{a}^{\lambda} dx \sqrt{2(V-E)}
\right \}} \ \lambda > a
\label{eq:wa}
\end{eqnarray}
Therefore, the particle density at the main well is given by
\begin{equation}
\rho (\lambda , E ) = \frac{1}{\sqrt{2(E - V(x))}} \cos^{2} \left ( N
\int_{0}^{x} d y \sqrt{2 ( E - V(y) )} \right ) .
\label{eq:c1}
\end{equation}
In the large $N$ limit the $cos^{2}$ must be replace by $1/2$ and the particle
density is given by the WKB formula\cite{amb}
\begin{equation}
\rho (\lambda , E ) = \frac{1}{\sqrt{2(E - V(\lambda))}}
\end{equation}
But, in the double scaling limit (\ref{eq:c1}) becomes
\begin{equation}
\rho ( x , e ) = \frac{1}{\sqrt{2(e - v(x))}} \cos^{2} \left (
\frac{1}{\hbar}
\int_{-\infty}^{x} d y \sqrt{2 ( e - v(y) )} \right )
\end{equation}
where $e$ and $v$ are the scaling part of the energy and the
potential, and $\hbar$ is the scaling coupling constant\cite{amb,mig}
\begin{equation}
\hbar^{2} = \frac{4 g_{c}^{\frac{5}{2}}}{N^{2} ( g - g_{c} )^{\frac{3}{2}}}
\end{equation}
where $g_{c}$ is the critical coupling constant.
The argument of the $cos^{2}$ is a
function of the scaling constant $\hbar$,
and now the particle density becomes
an oscillatory function if $\hbar$ is finite.
The amplitude is an increasing function of the
position and go to infinity when the position approach the
classical turning point because the WKB approach break down
at the turning points. In fact the
exact particle density is computed numerically in\cite{mig}
and it is given by an
oscillatory solution with an increasing amplitude when the position approach
the turning point, but, of course, is finite at the turning point. Therefore,
the oscillations of the particle density, which are nonperturbative in nature,
appears at the first WKB aproximation of the wave function.

The nonperturnative part is given by the same
oscillatory solution at the main well
with a phase sifted by $\pi / 2$, and an increasing exponential between the
cut $a$ and the local minimun $b$, and beyond the local minimun there is only
the perturbative wave function.
\begin{eqnarray}
\Psi^{np} (\lambda) & = & \frac{ - \tilde{A} N
}{ 2 \omega_{n} (2(E-V))^{\frac{1}{4}}}
\sin{ \left \{ N \int_{\lambda}^{a} dx \sqrt{2(E-V)} - \frac{\pi}{4} + \eta
\right \}} \ \lambda < a \nonumber \\
\Psi^{np}(\lambda) & = &
\frac{\tilde{A} f(e_{1},x_{0})}{(2(V-E))^{\frac{1}{4}}}
\exp{ \left \{ + N \int_{a}^{x} dx \sqrt{2(V-E)}
\right \}} \ a < \lambda < b
\end{eqnarray}
The amplitude of the nonperturbative
wave function is suppressed by a nonperturbative term
\begin{equation}
\tilde{A} \sim \exp \left\{ - 2 N \int_{a}^{b} d x \sqrt{2(V - E)}
\right\} A
\end{equation}

In reference\cite{mig} the particle density beyond the local minimun $b$
decrease very fast. In the double scaling limit the
perturbative wave function (\ref{eq:wa}) decrease as
\begin{equation}
\Psi \sim \exp \left\{ - \frac{1}{\hbar}
\int_{a}^{x} \sqrt{2(v - e )} \right\} \sim \exp \left\{
- \frac{1}{\hbar} x^{\frac{5}{2}} \right\}
\end{equation}
Between the cut $a$ and the local minimun $b$ the increasing exponential of
the nonperturbative wave function
gives some contribution to the particle density. Hence, the particle density
decrease more slowly between $a$ and $b$. However, the nonperturbative wave
function is supressed by a nonperturbative term, and the particle density also
decrease exponentially between $a$ and $b$. This is in agreement with\cite{mig}
where the particle density
decrease more slowly between the cut $a$ and the local minimun $b$ than beyond
the local minimun.

The above discussion suggest that there are only one perturbative vacuum,
given by $\Psi^{p}$, and
almost one nonperturbative vacuum, given by $\Psi^{np}$.
So the perturbative vacuum decay into
a nonperturbative vacuum. The nonperturbative vacuum break down the symmetries
of the perturbative one, so the Ward identities of the model (loop equation and
Virasoro constraints)
must change. Is an open question what kind of symmetries replaces the old one.

{\bf Acknowledgements}

I am grateful to J. Gonz\'alez for helpful discussions.

\newpage

{\bf Appendix 1}

In a neighbourhood of size $\sqrt{\hbar}$ around the local minimun
the Schr\"{o}dinger equation is given by
\begin{equation}
\frac{d^{2} \Psi}{d x^{2}} - ( e_{1} + C^{2} x^{2} ) \Psi = 0
\label{eq:a11}
\end{equation}
where $e_{1}$ is the first perturbative
correction to the Fermi level and is
positive because the perturbative
Fermi level is placed below the local
minimun. I have placed the origin of
coordinates at the local minimun of the
potential and $x = \lambda - b$.

Solutions of (\ref{eq:a11}) are given by
linear combinations of the even and odd
functions
\begin{eqnarray}
\Psi_{even} (x) & = & e^{-\frac{1}{2} C x^{2}} F ( \frac{1}{4} +
\frac{e_{1}}{4C} \mid \frac{1}{2} \mid C x^{2} )
\nonumber \\
\Psi_{odd} (x) & = & \sqrt{C} x e^{-\frac{1}{2} C x^{2}}
F ( \frac{3}{4} + \frac{e_{1}}{4C} \mid \frac{3}{2} \mid C x^{2} )
\label{eq:a12}
\end{eqnarray}
where $ F ( a \mid c \mid z ) $ is the Hipergeometric degenerate functions
and is the solution of\cite{table}
\begin{equation}
z F^{\prime \prime} + (c-z) F^{\prime} - a F = 0
\end{equation}

The asymptotic behaviour of (\ref{eq:a12}) can not match the WKB
wave function. However
for large values of $x$, $e_{1}$ is
small in comparison with the potential, and
one can approach (\ref{eq:a12}) by
Bessel functions of order $\frac{1}{4}$
at zero order in $e_{1}$. The
asymptotic behaviour of Bessel functions match the WKB wave functions.

Hence, I construct a wave function,
with the correct asymptotic behaviour and given
by (\ref{eq:a12}) around the local minimun, as follows.

The real axis is divided into three regions:
the neighbourhood of the local minimun, where the
wave function is given by Hipergeometric degenerate functions
\begin{equation}
\Psi_{1} = \alpha ( \Psi_{even} + \Psi_{odd} )
+ \beta ( \Psi_{even} - \Psi_{odd} ). \nonumber
\end{equation}
An intermediate
region, where the wave function is given by Bessel functions
\begin{eqnarray}
\frac{\Psi_{2}^{-}(-x)}{\sqrt{-x}} & = & A_{2}
\left [ I_{\frac{1}{4}} ( \frac{C x^{2}}{2} )
+ I_{\frac{-1}{4}} ( \frac{C x^{2}}{2} ) \right ] \nonumber \\
 & + & B_{2} \left [ I_{\frac{1}{4}} ( \frac{C x^{2}}{2} )
- I_{\frac{-1}{4}} ( \frac{C x^{2}}{2} ) \right ] \nonumber \\
\frac{\Psi_{2}^{+}(x)}{\sqrt{x}} & = & A_{1}
\left [ I_{\frac{1}{4}} ( \frac{C x^{2}}{2} )
+ I_{\frac{-1}{4}} ( \frac{C x^{2}}{2} ) \right ] \nonumber \\
 & + & B_{1} \left [ I_{\frac{1}{4}} ( \frac{C x^{2}}{2} )
- I_{\frac{-1}{4}} ( \frac{C x^{2}}{2} ) \right ]
\end{eqnarray}
The continuity of the wave function and its derivative is required in the
boundary $x_{0}$ of the above regions
\begin{eqnarray}
\Psi_{1} (x_{0}) & = & \Psi_{2}^{+} (x_{0}) \nonumber \\
\Psi_{1} (-x_{0}) & = & \Psi_{2}^{-} (-x_{0}) \nonumber \\
\left ( \frac{d \Psi_{1}}{dx} \right )_{ x = x_{0}} & = & \left (
\frac{d \Psi_{2}^{+}}{dx} \right )_ {x = x_{0}} \nonumber \\
\left ( \frac{d \Psi_{1}}{dx} \right )_{ x = - x_{0}} & = & \left (
\frac{d \Psi_{2}^{-}}{dx} \right )_{x = - x_{0}}
\label{eq:a13}
\end{eqnarray}
And the asymptotic
region, where the Bessel functions becomes WKB wave functions. The asymptotic
behaviour of Bessel functions are
\begin{eqnarray}
I_{\frac{1}{4}}(z) + I_{ - \frac{1}{4}}(z) & \rightarrow &
\frac{2}{\sqrt{2 \pi z}} e^{z} \nonumber \\
I_{\frac{1}{4}}(z) - I_{ - \frac{1}{4}}(z) & \rightarrow &
\frac{-2}{\sqrt{2 \pi z}} \sin ( \frac{\pi}{4} ) e^{-z}
\end{eqnarray}

The set of equations (\ref{eq:a13}) gives the connections formulas between the
asymptotic WKB wave functions.

The wave function must be the fall off exponential for $x \gg 0$. Hence,
$A_{1} = 0$, and $B_{1} = B$ is an arbitrary constant, which can be fixed by
normalization. For $x \ll 0$ the wave function is
\begin{eqnarray}
\Psi (\lambda) & = & \frac{ B f_{1}
(e_{1},x_{0})}{(2(V-E))^{\frac{1}{4}}} \exp {\left \{ - N \int^{0}_{x} dx
\sqrt{2(V-E)} \right \}} \nonumber \\
 & + & \frac{ B f_{2}
(e_{1},x_{0})}{(2(V-E))^{\frac{1}{4}}} \exp {\left \{ + N \int^{0}_{x} dx
\sqrt{2(V-E)} \right \}}
\label{eq:a14}
\end{eqnarray}
where $f_{1}$ and $f_{2}$ are given by the condition (\ref{eq:a13}).

I have constructed a wave function which is given by the usual WKB wave
function in the asymptotic region,
and in the interval $(-x_{0},x_{0})$, by the
exact solution of the aproximate Schr\"{o}dinger equation (\ref{eq:a11}).
However, this wave function is not unique because the point $x_{0}$
is arbitrary.

In the limit of $x_{0}$ small, the coefficients of the wave function
(\ref{eq:a14}) becomes
\begin{eqnarray}
f_{1}(x_{0}) & = & x_{0} \frac{1}{(\Gamma(5/4))^{2}} \sqrt{\frac{C}{4}}
+ O ( x_{0}^{2} )
\nonumber \\
f_{2}(x_{0}) & = & 1 + x_{0} \frac{\Gamma(3/4)}{\Gamma(5/4)}
\sqrt{\frac{C}{4}} + O ( x_{0}^{2} )
\end{eqnarray}

If one set $x_{0} = 0$, the wave function for $x \ll 0$ is given
only by the
increasing exponential. This connection condition also
arise from the Schr\"{o}dinger
equation (\ref{eq:a11}) if one set $e_{1} = 0$. Hence,
the decreasing exponential can
be interpreted as the first correction to the approach $e_{1} = 0$.

\newpage

{\bf Appendix 2}

In this appendix I show the proof of the Virasoro constraints
in the stochastic
stabilization of 2D quantum gravity. The proof of the Loop equation is
given in reference\cite{osc}.

Let be the matrix potential $W$
\begin{equation}
W = \sum_{n=0}^{\infty} \frac{2 g_{n}}{n} Tr \Phi^n
\end{equation}
the potential of the stabilized theory becomes now
\begin{equation}
V =  \frac{1}{2} \left \{ \sum_{n=0}^{\infty} \sum_{m=0}^{\infty} g_{n} g_{m}
Tr \Phi^{n+m-2} - \frac{1}{N} \sum_{n=0}^{\infty} g_{n} \sum_{p=0}^{n-2} Tr
\Phi^{p} Tr \Phi^{k-p-2} \right \}
\end{equation}
and from the Hellmann-Feynmam theorem
\begin{equation}
\frac{\partial E}{\partial g_{k}} = \frac{1}{2}
\left \{ 2 \sum_{n=0}^{\infty} g_{n}
\langle Tr \Phi^{n+k-2} \rangle - \frac{1}{N} \sum_{p=0}^{k-2}
\langle Tr \Phi^{p} Tr \Phi^{k-p-2} \rangle \right \}
\label{eq:a1}
\end{equation}

If the ground energy is zero, (\ref{eq:a1}) becomes the usual discrete
Virasoro constraints\cite{ito}
\begin{equation}
\left ( \sum_{n=0}^{\infty} g_{n} (k+n-2) \frac{\partial}{\partial g_{n+k-2}} +
\frac{1}{4N^{2}} \sum_{p=0}^{k-2} p(k-p-2) \frac{\partial^2}{\partial g_{p}
\partial g_{k-p-2}} \right ) Z = 0
\end{equation}

\newpage

{\bf Appendix  3}

In this appendix I perform the calculation of the first correction to the Fermi
level in the stabilized potential of the cuartic matrix model and I will show
that the Fermi level is placed below the local minimun.

The perturbative quantization condition for the Fermi level is given by
\begin{equation}
\frac{N}{\pi} \int_{-a}^{a} dx \sqrt{2(E_{F}-V)} - \frac{g}{2} \frac{1}{\pi}
\int_{-a}^{a}dx \frac{x^{2}}{\sqrt{2(E_{F}-V)}} + O(1/N) = N - \frac{1}{2}
\label{eq:a31}
\end{equation}
where the extra term arise from the
Hartree aproximation to the interacting term
(\ref{eq:A}). The Fock term is subleading in the quantization condition
(\ref{eq:a31}).

The first correction to the Fermi level is given by
\begin{equation}
E_{F}^{(1)} = \omega \left \{ - 1 + g \frac{1}{\pi} \int_{-a}^{a} d \lambda
\frac{\lambda^{2}}{\sqrt{2(E_{F}^{0}-V)}} \right \}
\end{equation}
where $\omega$ is the classical frecuency which is positive.

Following reference\cite{voz} $E_{F}^{(1)}$ becomes
\begin{equation}
\frac{E_{F}^{(1)}}{\omega} = - 1 + \frac{1}{\pi}
\frac{a^{2}}{2 b \sqrt{b^{2}-a^{2}}} \left \{ \pi - 2 \arctan{
\left (\frac{b}{\sqrt{b^{2}-a^{2}}} \right ) } \right \}
\end{equation}
and is easy to see that for
$b^{2} \geq a^{2} > 0 $, $E_{F}^{(1)}$ is negative and the
Fermi level is placed below the local minimun.

\vfill

\pagebreak

\end{document}